# Time synchronization in vehicular ad-hoc networks: A survey on theory and practice

Khondokar Fida Hasan, Charles Wang, Yanming Feng*, Yu-Chu Tian*

School of Electrical Engineering and Computer Science, Queensland University of Technology, GPO Box 2434, Brisbane QLD 4001, Australia

ABSTRACT

Time synchronization in communication networks provides a common time frame among all nodes, thus supporting various network functions such as message transmission, channel scheduling and resource sharing in real-time and in correct order. In vehicular ad-hoc networks (VANETs) for connected and automated vehicles, network nodes must be synchronized to exchange time-critical vehicle locations and warning messages for various road safety applications. However, this is challenging as the data communication systems need to achieve low latency and high reliability under the conditions of high user dynamics and density. While many synchronization techniques have been developed for general communications networks, it is necessary to understand the requirements for VANET time synchronization and the applicability of existing time synchronization techniques in VANET applications. This paper provides a survey on theory and practice of time synchronization in VANETs. It addresses some key factors in VANET time synchronization such as requirements analysis, precision, accuracy, availability, scalability and compatibility, and highlights the advantages of Global Navigation Satellite System (GNSS) in VANET time synchronization. Through this survey, some insights are developed into existing and emerging protocols for time synchronization in VANETs.

## 1. Introduction

There are various types of clocks for daily use, engineering and scientific purposes. Well-known examples include mechanic clocks, electric clocks and atomic clocks. Quartz clocks are electric clocks that keep time by counting oscillations of a vibrating quartz crystal. As the most widely used timekeeping technology in the world, they are used in most clocks and watches, as well as in computer and communication networks that keep time. Quartz clocks supplied by their manufacturers typically keep time with an error of a few seconds per week. Low-cost quartz movements are often specified to keep time within 1 second per day, i.e., 6 minutes per year. High accuracy is possible at a higher cost, but is also subject to the stability of the oscillator, particularly with change in temperature. Atomic clocks use an electron transition frequency in the electromagnetic spectrum of atoms as a frequency standard for their timekeeping element. They are the most accurate time and frequency standards known, but are alo too expensive for general computers and communication devices. Therefore, time or clock synchronization is required to maintain the same time among all low-cost or quartz clocks in a computer or communication network.

Time synchronization is a challenging task for a wireless communication network, especially for decentralized networks such as vehicular ad-hoc network (VANETs), where mobile nodes are vehicles travelling on roads. VANETs are developed by applying the principles of Mobile ad-hoc networks (MANETs) to vehicle domains. In comparison with general MANETs, VANETs have unique features such as a hybrid network architecture, dynamic topology, and time-sensitive applications. They are a specific type of ad-hoc networks that can be characterized by its intermittent connectivity and high network node speed [1,2]. Communications for various services in vehicular environments are highly reliant on the location and time information of network nodes. Timely delivery of various messages in a precise order is crucial for effective and efficient VANET services. Some VANET applications have a requirement of time offset tolerance below 100 ms. All these may become achievable when all network nodes operate on the same clock time.

As VANET is a distributed and decentralized network, VANET nodes are physically detached from each other. Thus, maintaining a network-wide single clock time is impossible for the whole VANET. This demands time synchronization services and applications among all network nodes. Time synchronization helps adjust the drifts of the clocks of all network nodes with respect to a global time standard or with each other. In this way, every node in the network can operate with the same notion of time. This supports reliable and precise time synchronization in various VANET services such as coordination, communication, security, and time-sensitive applications.

* Corresponding authors.
 E-mail addresses: y.feng@qut.edu.au (Y. Feng), y.tian@qut.edu.au (Y.-C. Tian).

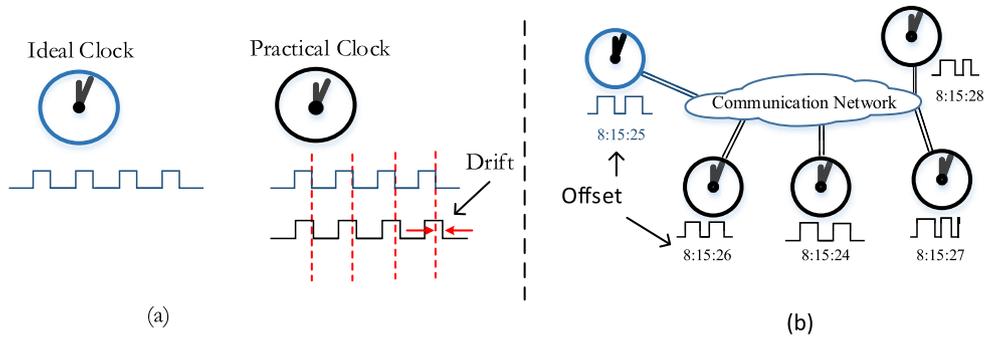

**Fig. 1.** Clock in a communication network. (a) Ideal and Practical physical clocks and their frequencies. (b) State of clocks in an unsynchronised communication network.

In general, merely adjusting the frequency of the clocks in a network refers to synchronization of frequency or syntonization. In contrast, synchronizing time means setting the clock to agree upon a particular epoch with respect to a standard time format such as Coordinated Universal Time (UTC). Synchronizing a clock refers to synchronization of both frequency and time. In this paper, the term *time synchronization* is used to refer to adjusting the clocks in the network to run at the same frequency. Therefore, the terms time synchronization and clock synchronization are used synonymously throughout this paper.

The overall aim of this paper is to provide a survey on the theory and practice of time synchronization in VANETs. The paper addresses the following three topics. Firstly, it presents the fundamentals of clock synchronization including challenges to achieve it in communication networks. Secondly, the paper provides a survey on the general approaches of time synchronization in wireless networks including a requirements analysis for time synchronization in VANET. From the requirements of time synchronization in VANET, a survey of existing approaches on time synchronization in VANET is also presented. Finally, the paper highlights the advantages of Global Navigation Satellite System (GNSS) for time synchronization in VANET and explains the basic principle of the operation of GNSS time synchronization in VANET.

## 2. Fundamentals of time synchronization in communication networks

Clocks used in communication networks may be grouped into hardware and software clocks. A physical hardware clock is made up of an oscillator to generate a pulse train and a counter to count and store the pulses. Hardware clocks can be constructed from different materials ranging from the most precise and expensive caesium, i.e., atomic clocks, to inexpensive quartz-powered clocks. A software or logical clock is a software-enabled programmable device that uses counting algorithms to track a local time value and maintain the time base of the system. Essentially in a standalone system, logical clocks follow a system's hardware clock. Thus, clock accuracy depends on the performance of the hardware clock.

The quality of a hardware clock, however, mainly depends on the stability of the oscillator and of the counting device. The stability is subject to changes of various parameters such as the nominal frequency of the oscillator, temperature, and other environmental factors. Such influences create a deviation in the device clock from the actual time. Such a deviation is known as clock drift. Fig. 1 shows the frequency of an ideal clock that is theoretically considered as a constant over time. However, in practice, the frequency changes due to both internal and external influences and drifts from its theoretical value. As a result, the time on each local clock system deviates from a more precise clock time and also from each other. This difference is known as time offset. Therefore, in a communication network as shown in Fig. 1(b), all the node clocks may report different times.

Operating communication networks requires alignment of node clocks to a reference clock, or synchronization of network time to a reference time. Fundamental operations may include successful communication, channel scheduling, real-time control messages. Alignment refers to reducing the effects of clock offset and drift between nodes to an acceptable level. A straightforward solution is to use an accurate source of time such as an atomic clock in every device of the network. However, this is too expensive and thus unrealistic in most real network scenarios. Communication network nodes are usually equipped with inexpensive quartz clocks. The technique of clock synchronization is used to equip all node clocks with the same time.

The basic idea of time synchronization is to minimize clock drifts and offsets resulting from various errors and inaccuracies. This is achieved by communicating messages that help transfer time from one node to another. Fig. 2 shows the fundamental concept of message transmission that synchronize clocks node by node. Ideally, such messages can be transmitted from a sender node to a receiver node or back and forth between the sender and the receiver to attain a common agreed time [3,4]. The accuracy and precision of clock synchronization, therefore, depend on the accurate transmission and reception of the messages. A number of synchronization protocols have been evolved targeting both wired and wireless networks over the time. In all cases, they deal with the fundamental problem of measuring the variation in sending and receiving time of messages, including access and propagation time over the medium by comparing the timing information received from the nodes [5,6]. The efficiency of synchronization protocols hence lies on the ability to accurately predict and eliminate message transmission-related delays by comparing their clocks.

From the fundamental concepts discussed above, the next section canvasses the technical details and principles of time keeping and clock synchronization.

## 3. Basic models and techniques of time keeping and time synchronization

This section presents general clock models and error sources that limit accurate time keeping towards achieving a common notion of time in a communication network. The levels of clock accuracy are also discussed together with basic techniques for time synchronization in decentralized communication network systems.

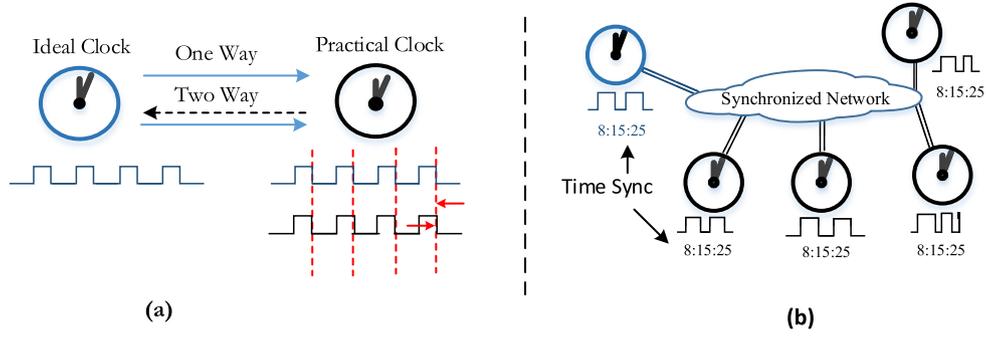

**Fig. 2.** Clock in a communication network. (a) Communication between an ideal (reference) clock and a practical node clock. (b) State of clocks in an synchronised communication network.

### 3.1. Hardware clocks

A hardware clock consists of a counter to count time ticks, which are ideally of a fixed length. A hardware oscillator updates the counter at a constant rate, i.e., frequency. The quality of the clock thus depends on the stability of the oscillator. Let $C(t)$ denote the reading of a clock counter $t$. Its rate of change at time $t$ is denoted by $f(t)$, i.e.,

$$f(t) = dC(t)/dt \qquad (1)$$

For an ideal clock, the rate is 1. However, a real clock fluctuates over time due to the fact that the rate changes because of various limiting factors. In a typical node $p$ with a quartz crystal, whose nominal frequency is defined as $f_p^0$, the relative frequency deviation is

$$\rho_p(t) = f_p(t)/f_p^0 - 1 \qquad (2)$$

According to [7], a model for a real clock is expressed as

$$f_p(t) = f_p^0 \cdot [1 + \rho_p^i(t) + \rho_p^a(t) + \rho_p^n(t) + \rho_p^e(t)] \qquad (3)$$

where, $\rho_p^i(t)$ is the initial frequency deviation at start time, $\rho_p^a(t)$ considers the aging effect, $\rho_p^n(t)$ denotes the jitter due to short-term noise, and $\rho_p^e(t)$ represents the jitter due to environment changes.

The environmental jitter $\rho_p^e(t)$ is a major factor influencing the quartz clock drift, in which variation in temperature typically contribute the most. According to [8], a jitter in order of $10^{-6}$ to $10^{-5}$ could be introduced by temperature changes. Other environmental influences in supply voltage and mechanical effects such as shock and vibration, can cause fluctuations. The short-term noise is typically in the order of $10^{-8}$ to $10^{-12}$. The aging effect $\rho_p^a(t)$ is in an order of $10^{-7}$ per month. Overall, the systematic deviation for the initial frequency while restarting an oscillator can grow at an order of $10^{-5}$ [9].

### 3.2. Software clocks

Software clocks or logical clocks are algorithms residing in programmable devices. They take local clock value $C(t)$ as input and convert it to time $S(C(t))$, which all programs use for time-dependent applications. This time $S(C(t))$ is the consequence of time synchronization. The mathematical model of such a typical software clock is

$$S(C(t)) = t_0 + C(t) - C(t_0) \qquad (4)$$

where $t_0$ is the (correct) real time. Such a software clock runs with the speed of the hardware clock.

### 3.3. Clock accuracy and precision

*Clock accuracy* and *Clock precision* are two related yet different concepts. Clock accuracy, denoted by $\alpha$, refers to the degree of correctness of the clock time. In comparison, clock precision refers to the consistence of the clock time with some other and/or standard clock. In synchronization nomenclature, the accuracy is the largest or maximum acceptable clock offset between the node clock and the reference clock. It is determined by measuring the mean of the error between the node and external reference clock and usually represents the synchronization bias [10]. The clock of a Node $p$ can run with the accuracy $\alpha$ if the clock value $C_p(t)$ is in an open $\alpha$-neighbourhood around the standard absolute time $t$ in an observable period of $T$ [7,11,12], thus,

$$|C_p(t) - t| \leq \alpha, \ \forall t_a T \qquad (5)$$

The clock *precision* $\beta$, however, is the measure of the standard deviation of the mean clock error and quantifies the synchronization jitter. It is often also called *instantaneous precision*, which represents the boundary of the difference between two clocks $p$ and $q$, i.e.,

$$|C_p(t) - C_q(t)| \leq \beta, \ \forall t_a T \qquad (6)$$

In internal synchronization environments, the clock precision is the maximum difference between two clocks. For external synchronization with a standard time, this difference is the accuracy as expressed in Equation (5).

### 3.4. Clock offset, skew and drift

A free-running clock is influenced by a number of factors as described in Equation (3). It fluctuates and deviates from the actual time due to the clock drift. Theoretically, software clocks are similar to hardware clocks, as software clock algorithms follow the system clocks, which depend on hardware clocks. Thus, the accuracy of software clocks relies on the accuracy of hardware clocks.

The accuracy of a clock as defined above pertains to the overall degree of clock uncertainties relative to a reference standard time. Clock uncertainties can be further described through *Offset*, *Skew* and *Drift* [13]. *Clock Offset* is defined as the time differences between a clock time and the standard true time. It is $|C_p(t) - t|$ for node $N_p$. It is seen from Equation (5) that the clock accuracy is the absolute value of the clock offset. The relative clock offset between two nodes $N_p$ and $N_q$ at time $t$ is expressed as

$$\text{Clock Offset} = C_p(t) - C_q(t) \qquad (7)$$

*Clock Skew* is defined as the difference of the clock frequencies between a system clock and a perfect clock. It is the first derivative of the clock offset with respect to the real time $t$. The clock skew of a clock $C_p$ relative to $C_q$ at time $t$ can be expressed as

$$\text{Clock Skew} = C'_p(t) - C'_q(t) \qquad (8)$$

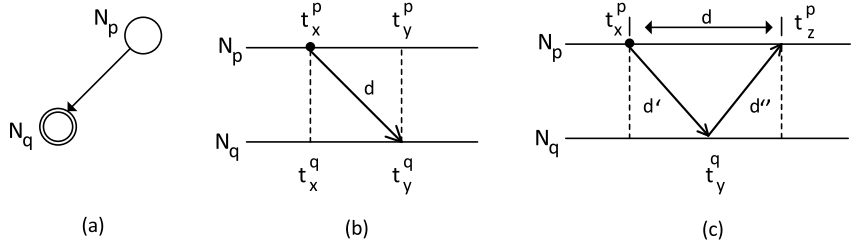

Fig. 3. Message exchanges between two nodes.

*Clock Drift* of a clock $C_p$ is defined as the second derivative of the clock value with respect to the real time $t$, i.e., $C''(t)$. Therefore, the relative clock drift between two nodes $N_p$ and $N_q$ is represented by

$$\text{Clock Drift} = C''_p(t) - C''_q(t) \qquad (9)$$

Overall, the above three terms are frequently used to characterize the performances of a typical clock in a communication system.

### 3.5. Main limiting factors in time synchronization

The performance of time synchronization methods is affected by two main factors: the inherent performance of the clock oscillators and how effectively a chosen synchronization technique works between them. The systematic and random errors of clock oscillators accumulate over time [14], which impact on the synchronization accuracy.

Several issues in synchronization techniques can affect the clock synchronization. The first issue is the capability of the technique to deal with the uncertainty of message delay during radio communication. Other issues include *Clock Adjustment Principle* and *Timestamping Accuracy*. The estimation of various latencies during *Sending time*, *Accessing time*, *Propagation time* and *Receiving time* is crucial for adjusting clocks precisely over a network. The clock adjustment performance is highly dependent on the method and quality of the synchronization algorithm. *Timestamping* is a method of adding time into the packet during the transmission and reception of a message. In packet-based synchronization techniques, precise time-stamping is crucial. By calculating the egress and ingress timestamps, the propagation delay is measured. It is known that the accuracy of clock synchronization varies from one protocol layer to another [15–17]. This is due to the uncertainty of inter-layer delays. Physical layer time-stamping is considered to be the most accurate way so far. After receiving the timestamps, a node needs to run an operation to adjust the clock. The performance of this adjustment operation also determines the accuracy and quality of the synchronization technique.

### 3.6. Basic techniques for time synchronization in a decentralized system

VANETs are decentralized systems. To explain the time synchronization mechanism in a distributed or decentralized network, we consider a model involving two nodes $N_p$ and $N_q$, as shown in Fig. 3. When Node $N_p$ sends a message with its local time stamp $t_x^p$ to Node $N_q$ (Fig. 3(b)), Node $N_q$ receives the signal at $t_y^q$ and updates its time accordingly. This is known as *Unidirectional Synchronization*. In unidirectional synchronization, the transmission delay is not considered. It suffers from a large synchronization error. Therefore, a more complicated *Round-Trip Synchronization* tech-nique is more acceptable. In this technique, Node $N_p$ sends message at $t_x^p$ to node $N_q$ to ask for the timestamp $t_y^q$. After getting the response from node $N_q$, node $N_p$ performs calculation to determine the round-trip time $d = t_z^p - t_x^p$. This round-trip time is basically the time interval of two-way message transmissions as shown in Fig. 3(c). Then, it is used to improve the precision of the time synchronization between the two nodes. The drawback of this synchronization method is the introduction of message exchange overheads.

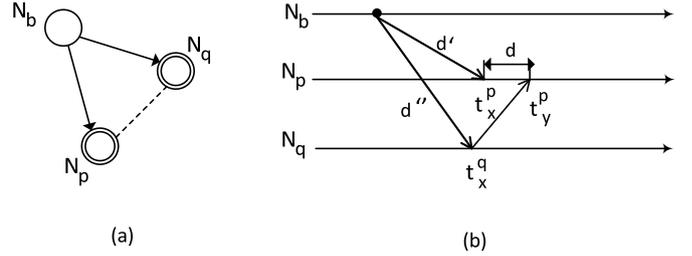

Fig. 4. Reference Broadcasting Synchronization (RBS).

Another effective method, namely, packet-based clock synchronization is *Reference Broadcasting Synchronization* (RBS). Its operation is shown in Fig. 4. In RBS, a beacon sends a synchronising message to all nodes. For example, in Fig. 4, node $N_b$ is the beacon node. It sends beacon a message to nodes $N_p$ and $N_q$. The delay $d'$ for $N_p$ and delay $d''$ for $N_q$ are almost the same. After receiving the beacon signal, $N_q$ sends its time stamp $t_x^q$ to node $N_p$. Then, node $N_p$ calculates the time interval $d = t_y^p - t_x^p$. The result is a measure of the time difference between nodes $N_p$ and $N_q$.

### 3.7. Types of clock synchronization

Several parameters such as the source of the reference clock, the required accuracy of the synchronization, the communication medium between nodes and the supported applications can all impact on the method of clock synchronization used. Therefore, depending on the variation of methods and their applications, clock synchronization can be classified differently.

For example, when a system maintains synchronization with a standard reference clock time, it is known as absolute clock synchronization. When the nodes in a network are synchronized with each other with respect to time, the method is known as relative clock synchronization [18]. From the variation of the synchronization protocols, we can classify time synchronization differently. Some time synchronization protocols commonly differ from each other in some aspects but sometimes resemble each other in some other aspects [13]. For instance, consider deterministic and probabilistic clock synchronization. Deterministic protocols stipulate a strict upper bound on the offset error certainty compared to probabilistic synchronization where it uses fewer message transfers and, therefore, less processing overhead [19,20].

However, the most popular way of classifying clock synchronization methods is based on the time references system that is used. According to the time scale, clock synchronization in dis-tributed network can be classified into two main types: synchronization with internal time-scale and synchronization with external time-scale. In ad-hoc networks such as Wireless Sensor Networks (WSNs) and VANET, time synchronization can be implemented locally with an internally consistent time-scale. However, VANETs are outdoor wireless ad-hoc networks. Therefore, it is also possible to deploy global time synchronization with an external time scale.

Synchronization methods with an internal time-scale is realized through a set of operations and message exchanges between nodes. This requires estimating both offset and skew of the local clocks relative to each other. Hence, synchronization with internal time-scale maintains a relative time notion with respect to each other. Such relative synchronization is the basis of most indoor networks such as indoor wireless sensor networks and Wi-Fi.

Synchronization methods with an external time-scale method are implemented with respect to an absolute or external reference time standard, such as UTC. Such an external reference time-scale is usually transmitted and distributed by using a global radio system. Typical global radio systems include satellite-based GNSS and the short-wave WWVB station [21,22].

The next sections examine the synchronization techniques currently practiced in different wireless networks.

## 4. Approaches to the time synchronization in wireless media

Most time synchronization protocols for communication networks are applicable in both wired and wireless media. For example, the well adopted Network Time Protocol (NTP), which is considered as the backbone of wired computer communication networks, appears to be implementable in wireless media with certain accuracy [23,24]. However, based on the similar principles, the performance is yet to be improved in further technological evolutions.

This section explores the prominent synchronization techniques over WSN. This provides a basis for synchronization in vehicular wireless networks. This is followed by a presentation of the challenging issues and requirements of time synchronization in vehicular networks.

### 4.1. Revisiting time synchronization in wireless sensor networks

As a VANET is a type of mobile ad-hoc wireless networks, it is worth examining the existing synchronization techniques for other types of MANETs. The most straightforward one is WSNs, for which considerable research efforts have been directed to time synchronization. Five main WSN synchronization techniques are to be discussed below: time-stamp synchronization (TSS), reference-broadcast synchronization (RBS), lightweight time synchronization (LTS), a timing-sync protocol for sensor networks (TPSN), and flooding time synchronization protocol (FTSP).

*Time-stamp Synchronization (TSS)* is a WSN time synchronization method based on internal synchronization on demand [25]. TSS does not use specific synchronization messages for time synchronization. Instead, it uses timestamps embedded in other packets to perform synchronization post-facto. The time offset is estimated through calculation of the round-trip delay between the transmitters and receivers. For single-hop WSNs, the average uncertainty of TSS is recorded as 200 μs. In multi-hop networks, the maximum uncertainty of 3 ms is achieved in 5 hops.

*Reference-Broadcast Synchronization (RBS)* uses beacon broadcast for time synchronization. In RBS, any nodes in a basic single-hop network can send a beacon to broadcast its time reference. A node compares its local reference time with the reference times received from other neighbour nodes and adjusts its clock accordingly. RBS performs both offset and rate corrections when updating the clock. Making use of physical layer broadcasts, it does not carry explicit time-stamps. This synchronization is enacted for the whole network.

In a multi-hop network, all network nodes are grouped into clusters. In each cluster, a single beacon is used to synchronize all nodes in the cluster. A gateway node is used to transfer time-stamps from one cluster to another. This helps maintain the same reference time to compute offset and rate corrections. RBS uses the last minute time-stamps in order to reduce random hardware delay and access delay. Its average uncertainty is measured as 11 μs in laboratory experiments with 30 broadcasts. For multi-hop networks with $n$ hops, the average error grows in $O(\sqrt{n})$. While RBS provides comparatively high accuracy, it is subject to excessive protocol overheads.

*Lightweight Time Synchronization (LTS)* aims to reduce the complexity of synchronization overhead [26]. Therefore, unlike other synchronization methods, it provides synchronization with a specified precision. As a centralized algorithm, it begins with the construction of a spanning tree for the network with $n$ nodes. Next, a pair-wise synchronization is performed along the $n-1$ edges of the spanning tree. The root of the spanning tree works as the reference node. It initiates all on-demand resynchronization operations. The average synchronization error in LTS is recorded as 0.4 s. The maximum error can reach as high as 0.5 s.

*Timing-Sync Protocol for Sensor Networks (TPSN)* is a network-wide synchronization protocol based on a hierarchical approach [27]. It follows the classical approach of sender–receiver synchronization to create a hierarchical topology. The hierarchy maintains multiple levels in order to distinguish nodes to perform actions. TPSN performs time synchronization through two phases. In the first phase, a node at level 0 acts as the root node. It initiates a 'level discovery' broadcast message with its identity and level in the hierarchy. Its immediate neighbours receive this message and assign themselves level 1 below the root node. After that, each node at level 1 broadcasts a 'level discovery' message, which will be received by other neighbour nodes at lower levels. This process continues until all nodes are reached by such 'level discovery' messages. In the second phase, all nodes synchronize their clocks to their root or parent nodes in the tree by using a round-trip synchronization operation. This round-trip synchronization is conducted at the MAC layer. Therefore, message-delay uncertainties are largely eliminated. The accuracy of TPSN is considerably high. Experimental results show that TPSN synchronization of two Berkeley motes has reached an accuracy of 17 μs. A drawback of TPSN is the significant message exchange overhead particularly for a large number of nodes.

*Flooding Time-Synchronization Protocol (FTSP)* is a hybrid time synchronization protocol built upon RBS and TPSN [28]. In FTSP, a node with the lowest node identity becomes the root node, which works as the reference time sender. If this node fails, a node with the next lowest identity becomes the root node. The root periodically floods the network with synchronization message with the reference time. In this way, the whole network becomes synchronized. FTSP is a self-organized algorithm. It constructs a hierarchy to perform low-level time stamping and local clock correction. A FTSP experiment with an eight-by-eight grid of Berkley motes shows an average error of 1.7 μs with the maximum of 38 μs per hop.

### 4.2. Why is time synchronization an issue in VANETs

Time synchronization in distributed network systems is a well-recognized problem. In wireless communication networks, time synchronization is considered as a key element for consistent data traffic and also for accurate real-time control of message exchanges [29]. Many network applications require precise clock synchro-nization among the nodes to ensure correctly ordered operations. Otherwise, the performance of these applications and hence the network operations could be disrupted. Over the years, the issue of time synchronization has been extensively investigated in the context of computer and telecommunication networks [30–32]. Many protocols have been proposed and implemented to perform time synchronization over computer and telecommunication networks. Those protocols vary in terms of the required precision of timing and also according to the services and types of networks. For example, in a routed network, physical time is not a critical issue. Thus, protocols based on a routed network, i.e., Packet-over-SONET/SDH links (POS), requires synchronization to ensure the sequence of the order. On the other hand, some networks require synchronization with high time accuracy [33–35]. For example, in pure synchronous optical networking (SONET) and synchronous digital hierarchy (SDH) networks, the precision of time along with fixed time-division multiplexing mechanism is mandatory.

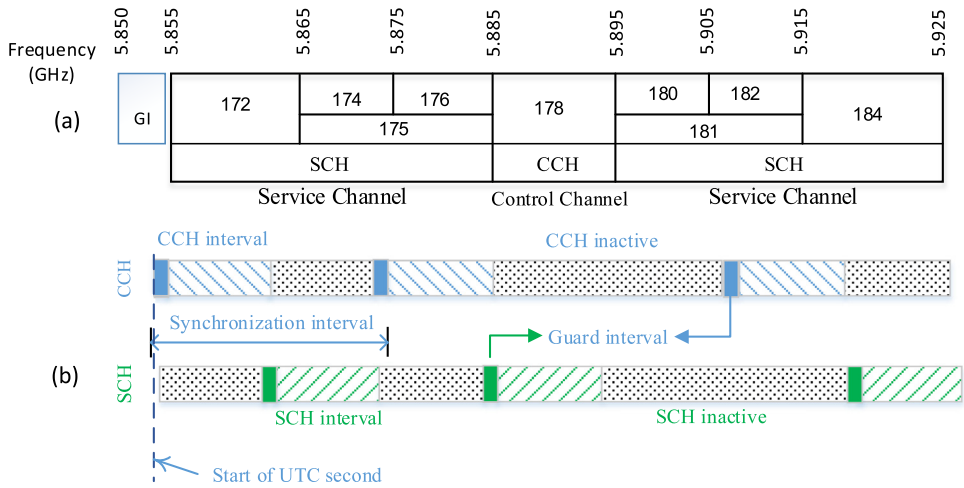

**Fig. 5.** WAVE spectrum: (a) Frequency and channel allocation; and (b) channel synchronization and guard interval.

In VANET, physical time plays an important role in many applications, which cannot be satisfied by logical time or any kind of event ordering models. Most communicating interactions for time-based decisions rely upon a time-of-day clock. For example, VANET enables traffic management on individual levels by providing communication among vehicular nodes and share road information such as vehicle dynamics, and driving intentions [36–38]. The current status of the nodes in a VANET, therefore, needs to be determined precisely in terms of position, speed and other real-time values. This frequently scheduled work requires time synchronization to develop accurate and precise time on node.

VANETs increase road safety by enabling different critical safety applications. Forward Collision Warning (FCW), Cooperative Collision Warning (CCW), Emergency Electronic Brake Lights (EEBL) are a few examples that alert a driver about possible crash scenarios ahead. In these applications, each vehicle is required to broadcast their basic safety messages (BSM) including vehicle location states periodically at 10 Hz. Event trigging warning messages are time sensitive and need to be transmitted and received orderly securing stringent delay requirements (typically 100 ms [39]). If nodes clock in VANET does not have any commonly agreed accurate time maintained among them, such periodical and event trigging safety message from the sender may report with a past timestamped information or with advanced timestamped information with respect to the receiver time. In either case, those messages may be discarded or mixed up after reception by the receiver nodes considering as an outdated message. Under such circumstances, a warning message would fail to alert drivers, thus leading to a risk of avoiding collision or road casualties. Time synchronization in VANET, therefore, is essential to achieve accurate and precise time over the network [40].

Physical time is also crucial for proper bandwidth utilization and efficient channel scheduling. Therefore, it is required that all the nodes in a VANET are able to report the same time, regardless of the errors of their clocks or the network latency the network nodes may have.

Furthermore, certain security measures in VANETs, such as duplication detection and identification of session hijacking and jamming, require absolute time synchronization [41,42]. Time plays a critical role in determination of two distinct real-world events to develop traceable communication for reconstruction of packet sequence on the channel [43].

### 4.3. Requirements analysis for time synchronization in VANET

Time synchronization assures that all nodes in a network are equipped with the same reference clock time. Maintaining a common notion of time is necessary in VANETs for various applications and also many system-level protocols. Therefore, time synchronization requirements in VANET involve not only the synchronization accuracy but also the performance, compatibility and feasibility characteristics of the synchronization. This section analyses the time synchronization requirements in VANETs.

The set of requirements for time synchronization in VANETs can be classified into two categories: *Performance-oriented requirements* and *Application-oriented requirements*. The former is based on the system-wide objectives including protocol support and compatibility with the synchronization method, whereas the latter is based on the needs of end applications.

#### 4.3.1. Performance-oriented requirements

Better spectrum utilization is considered as one of the key targets in wireless communication. VANETs use a limited spectrum, for instance, 75 MHz at the 5.9 GHz band. Better use of this spectrum enables to increase the capability of network bandwidth and throughput. In VANET protocol stack "Wireless Access for Vehicular Environment (WAVE)", the available bandwidth is divided into service and control channels, as shown in Fig. 5 [44–46].

For efficient channel coordination, communicating nodes need to be synchronized. In practical operations, the clocks of all nodes are delayed for many reasons and may tend to lose their synchronization. To accommodate the time differences among the nodes, a Guard Interval, which is also known as a Guard Band, is used in communication design. As shown in Fig. 1(b), a guard Interval is a period of time for separation of two consecutive and distinct data transmissions from different users in a time slotted mechanism or from the same users in a frequency slotted mechanism.

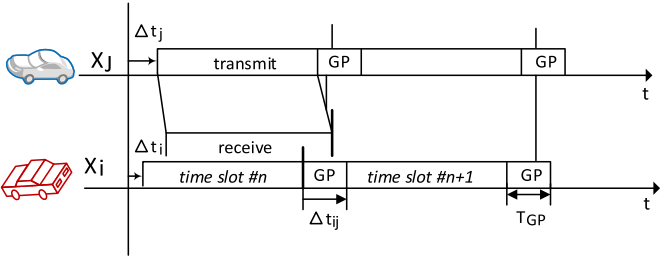

**Fig. 6.** Guard interval requirements.

The requirements of Guard Interval and its relationship with time synchronization accuracy in VANETs are graphically demonstrated in Fig. 6. As shown in Fig. 6, assume that nodes $N_i$ and $N_j$ have time offsets of $\Delta t_i$ and $\Delta t_j$, respectively, with reference to the global standard of time. When node $N_j$ sends a burst to node $N_i$, then the observed time offset $\Delta t_{ij}$ at node $N_i$ is estimated as

$$\Delta t_{ij} = \Delta t_j - \Delta t_i + d_{ij}/c \quad (10)$$

where $d_{ij}$ is the radio propagation distance between the two nodes, and $c$ is the speed of light, respectively. A successful reception of data at node $N_i$ from node $N_j$ can be achieved if there is no any overlap of communications due to time offsets. Therefore, a guard interval ($T_{GP}$) is introduced to avoid such an overlap. This requires the guard interval $T_{GP}$ to be greater than the time offset $\Delta t_{ij}$, i.e.,

$$\Delta t_{ij} < T_{GP} \quad (11)$$

Equation (11) is required for time synchronization in wireless networks. It is seen from this requirement that the guard interval can be reduced if VANETs are better time synchronized. Moreover, since guard interval is an addition to the communicating slot length, it consumes spectrum resources and consequently leads to a longer time to transmit a message. Therefore, a reduced guard interval implies an increased spectrum utilization.

The impact of the guard interval on the performance of VANET services varies with the type of the underlying communication protocol. Asynchronous wireless communication protocols in IEEE 802.11 networks use Carrier-sense multiple access with collision avoidance (CSMA/CA) as the channel coordination mechanism, in which a Guard Interval is used to avoid transmission disruption due to propagation delays, echoes and data reflections. In 802.11n networks, cutting the guard interval by half from 800 ns to 400 ns leads to an increase in effective data transmission rate by 11% [47].

In synchronous slotted protocols, e.g., Time-division multiple access (TDMA) and Space–time division multiple access (STDMA), a guard interval accommodates clock inaccuracies. This enables to avoid message collisions and message losses in time slotted medium access protocols. For example, the commonly used frame length of STDMA in Automatic Identification System (AIS) of ship navigation is 2,016 slots. In such a framing, a reduction of 10 µs in guard interval means to accommodate 45 new slots for every 496 µs, thus increasing the channel capacity by about 9% [48]. Therefore, precise time synchronization contributes to increased communication capacity of wireless networks.

In VANETs, network environments change frequently over time. One scenario of environment changes is dynamic changes in the network density from a small number of nodes ($<20$) to a large number of nodes ($>100$). To ensure the Quality of Service (QoS) in response to these changes in network density, an efficient, reliable and scalable medium access control mechanism is required with precise time synchronization. Another scenario of environment changes in VANETs is dynamic changes in location of network nodes from one geographical region to another, implying high mobility. Due to their fast moving nature, nodes connect and disconnect momentarily from the clusters. Maintenance of QoS in highly mobile networks requires all nodes to follow the same time standard, which is achieved through time synchronization.

*4.3.2. Application-oriented requirements*

In comparison with many other wireless ad-hoc networks, VANET nodes are noticeably dynamic and highly mobile. In VANETs, the relative speed between two nodes can be as high as over 200 km/h. This implies that a node with such a high speed will stay only a few seconds within the transmission range of other nodes. Moreover, some VANET applications require an extremely small end-to-end delay. For example, the maximum acceptable end-to-end latency for pre-crash sensing warning messages is 50 ms. In a recent work [49], the lowest end-to-end latency requirement is identified for cooperative sensing is 3 ms. For Lane Change Warning and Forward Collision Warning, it is specified to be 100 ms by National Highway Traffic Safety Administration of US [39,50,51]. Therefore, to meet the requirements of these safety applications, accurate timing is required with deterministic and reliable communications in VANETs.

Secure communications are also a key challenge in deployment of VANETs. Some threats such as session hijacking and jamming are identified to be dangerous in VANETs. Reconstruction of channel activity with traceable and reliable communications is considered being the preventive and forensic measures of such threats. The effectiveness and accuracy of activity reconstruction depends on the precision of time synchronization. Fine-gained analysis of channel activity between concurrent transmissions requires stringent synchronization guarantees of 8 µs for Dedicated Short-Range Communication (DSRC) technology [43].

## 5. Advances towards VANET time synchronization

Each of the wireless nodes in a VANET has a clock to keep time independently. This time can be local time or universal time and needs to be synchronized. This section begins by presenting a vehicular ad hoc network communication model to aid understanding time synchronization methods. Then recently proposed and currently practiced time synchronization methods in vehicular networks are examined.

*5.1. Communication model*

Communications in VANETs are governed by the protocol stack WAVE, which conforms to a series of IEEE standards. The WAVE controls the wireless medium for VANET through a bundle of IEEE protocols, such as IEEE 1609 (e.g., 1609.2, 1609.3, 1609.4) and IEEE 802.11p. IEEE 802.11p is an amendment to IEEE 802.11 for regulation of data link and physical layers.

The basic communication architecture in VANETs consists of two blocks: On Board Units (OBU) and Road Side Units (RSU). An OBU is a vehicle, while an RSU refers to the road side infrastructure for communications [39,52,53]. This is illustrated in Fig. 7.

*Vehicle-to-Vehicle* (V2V) communications are typical ad-hoc network communications. In V2V, mobile nodes communicate directly with each other. In such communications, all types of packet deliveries, such as unicast, multicast and broadcast, take place between vehicles without the intervention or support of any other network components.

*Vehicle-to-Infrastructure* (V2I) communications are implemented through wireless interactions between OBUs and RSUs. They enable real-time services, such as traffic information and weather updates.

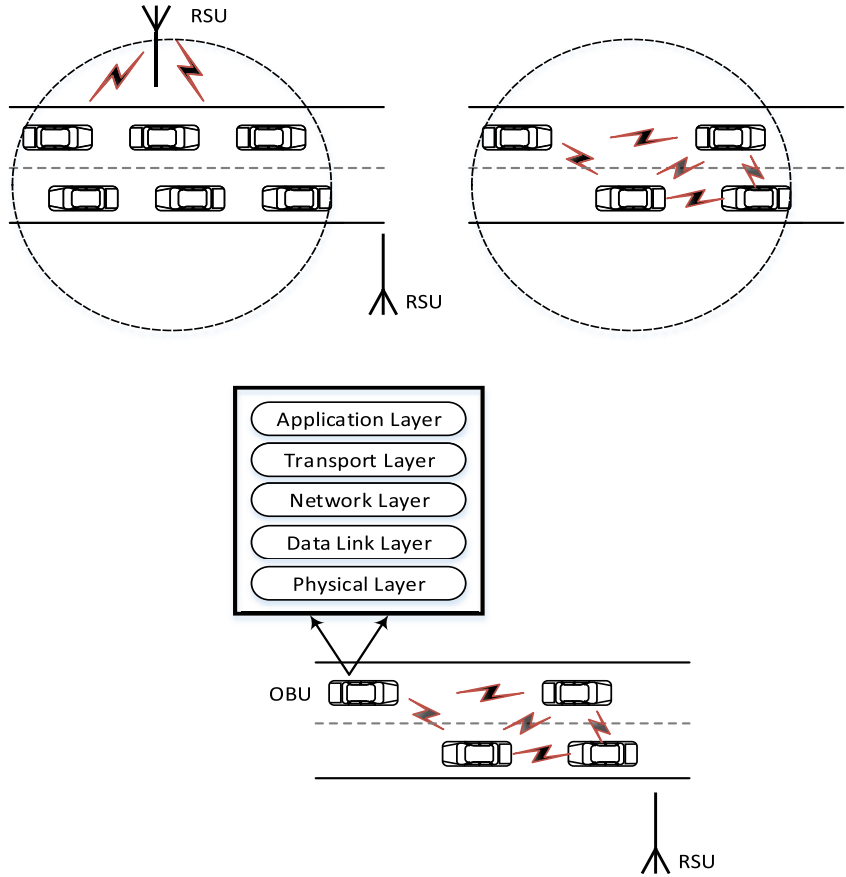

Fig. 7. VANET communication architecture.

V2I also provides support for secure sparse and long-distance communications [54].

In addition, *Internet infrastructure*, *private infrastructure* and *in-vehicle* communications also support some VANET services and applications. This can assist with remote identification of vehicle's performance and monitoring drivers' conditions such as fatigue and drowsiness. *In-vehicle* communications are considered as an architectural part of the latest definition of VANET communications. It is a significant component of safety and other applications in VANETs.

### 5.2. Existing recommendation for VANET time synchronization

Time synchronization for VANET has been solely based on protocol IEEE 802.11p. IEEE 802.11p is an amendment to Wireless Local Area Network (WLAN) protocol IEEE 802.11. Therefore, the synchronization technique from IEEE 802.11 family is naturally applicable to VANETs. In IEEE 802.11 standard family, a station (STA) can be attached to an Access Point (AP) in a centralised mode called Basic Service Set (BSS). It can also communicate with other STAs in decentralised ad-hoc mode called Independent BSS (IBSS). These two modes are shown in Fig. 8.

In 802.11 networks, time synchronization is predominantly required for frequency hopping and scheduling of sleep phases. The standard requirement of time synchronization is 274 μs, which is also the threshold of out-of-synchronization [55]. Time synchronization within 802.11 systems rely on a Timing Synchronization Function (TSF) timer. The TSF timer is a 64-bit hardware counter with a resolution of 1 μs and thus is capable of performing $2^{64}$ modulus counting. It employs a local clock oscillator built on WLAN chipset with a frequency accuracy of ±0.01%. The adjustment of the timer and hence the accuracy of synchronization depends on the operation mode, e.g., BSS mode in centralised communications or IBSS mode in a decentralised network. In BSS, the AP transmits beacons with TSF timer values, and an STA sets its own TSF timer with delay usually corrected by offset adjustment without rate correction.

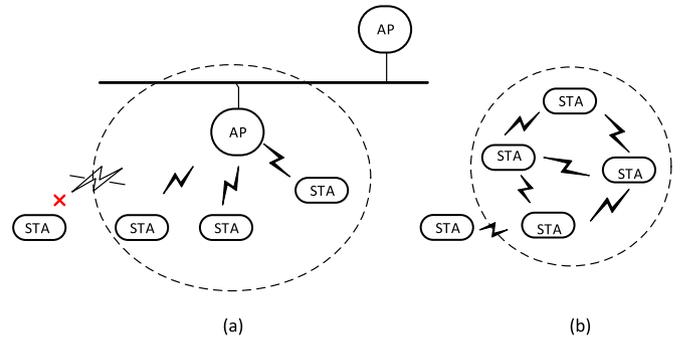

Fig. 8. Two modes of communications in 802.11 standard family.

In the infrastructure-based BSS mode, an AP acts as a master clock. It broadcasts the reference time for all STAs to be time synchronized. When beacon transmits, other data exchange operations are suspended so that the master can broadcast TSF synchronization values to all attached STAs. The period of beacon transmission depends on the network resource sharing mode as shown in the Table 1. In this mode, the receiving station only accepts TSF values and updates its clock.

In the ad-hoc IBSS mode, all STAs adopt a common value, aBeaconPeriod, which characterises the length of a beacon interval. At the beginning of the beacon interval, a beacon generation window forms. It consists of $\omega + 1$ as shown in Fig. 9. For the station that initiates IBSS, this interval also defines Target Beacon Transmission Times (TBTTs) in aBeaconPeriod times apart. A time zero is defined to be a TBTT. At the TBTT event, all STAs perform the following process:

1. At TBTT, suspend the backoff timer for any pending non-beacon transmission. The STA calculates a random delay distributed in the range [0, $\omega$), where $\omega = 2*$aCWmin$*$aSlotTime.
2. All STAs wait for the period of the random delay.
3. If the beacon is received before the expiration of the random delay timer, cancel the remaining random delay.
4. When the random delay timer expires, STA sends beacon using its TSF timer value as a timestamp.
5. When a station receives the beacon, it updates its TSF timer following the timestamp of the beacon if the beacon value is later than the station's TSF timer.

**Table 1**
Slot time with beacon generation window.

|                  | FHSS |   | DSSS |   |     |    | OFDM |    |    |
|------------------|------|---|------|---|-----|----|------|----|----|
| aCWmin           | 15   |   | 31   |   |     |    | 15   |    |    |
| aSlotTime (μs)   | 50   |   | 20   |   |     |    | 9    |    |    |
| Speed (Mbps)     | 1    | 2 | 1    | 2 | 5.5 | 11 | 6    | 12 | 24 | 54 |
| Beacon length    | 13   | 8 | 34   | 22| 14  | 12 | 12   | 8  | 5  | 4  |

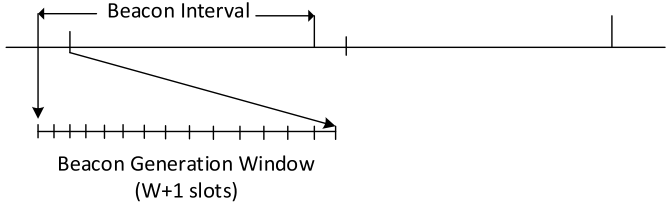

**Fig. 9.** Beacon generation window.

Therefore, the TSF synchronizes timers with the fastest STA in IBSS.

The above synchronization procedure is designed for single-hop networks. Such a procedure with TSF suffers from poor scalability and inability to handle congestions. When the number of nodes increases, the node with the fastest clock faces difficulties in successfully sending out beacon frames. As a result, its clock gradually drifts away from the clocks of other nodes. This problem is known as *fastest node asynchronism* [56].

In a multi-hop network employing the native IEEE 802.11 clock synchronization mechanism, the whole network is partitioned into multiple disjoint clock islands. If every island is out of synchronization with one another, the *time partitioning* problem appears [57].

Improved techniques have been proposed to address the so-called fastest node asynchronism problem and the time partitioning problem. The basic ideas are to enhance scalability and mitigate congestion. Two well-recognised improvements are Adaptive TSF (ATSF) and Multi-hop TSF (MTSF).

ATSF modifies the basic 802.11 TSF. It adds a priority scheme to overcome the *fastest node asynchronism* problem. This method involves maintaining and adjusting the transmission frequency of the beacon [56]. When a node receives a beacon message with a larger timestamp, it reduces its beacon transmission frequency. It keeps updating the beacon transmission frequency until it reaches the maximum allowed value. This allows the fastest node to have a higher probability of transmitting beacon messages.

In MTSF, each node maintains a path to the fastest node. The beacon is transmitted from the fastest node to all other nodes without being suppressed anywhere in the middle of the network. MTSF consists of two phases: a beacon window phase and a synchronization phase [58]. In the beacon window phase, all neighbour nodes construct a synchronization group and identify the fastest node as the root node of the group. In the synchronization phase, root nodes are synchronized with each other. In this way, the fastest node asynchronism problem can be avoided together with the partitioning problem.

The average maximum clock drift with TSF is 124.5 μs for 20 nodes. It increases to 500.2 μs when the number of nodes is 60. In comparison, MTSP performs much better. Experimental measurements show that the average clock accuracy of MTSF is 22.4 μs for 20 nodes and 39.1 μs for 60 nodes, respectively [59].

Such TSF-based synchronization lacks support from timing standards such as UTC, TAI etc. The 2012 amendment of IEEE 802.11 proposes two techniques, i.e., Timing Advertisement (TA) and Timing Measurement (TM) mechanisms, to obtain the support of global time [60]. In TA, the external reference clocks are attached to Access Points (APs). In TM, the frames use physical layer timestamps to perform synchronization between AP and STA, thus reduces multi-hop errors. However, the TA mechanism architecture requires a cascade of four clocks, which does not direct how the external clock will be synchronized to the AP and perform accurate time-stamping. References [61] and [10] have discussed details on that issue and proposed some measures in WLAN scenarios. VANET networks are more ad-hoc in nature compared to WLAN, where a large portion of it relies on STA to STA but STA to AP communication. Therefore, the feasibility of employing such a mechanism in VANET requires an extensive investigation.

## 6. GNSS approaches for VANET time synchronization

GNSS is a well-established international utility for positioning, navigation and timing (PNT). The generic term "GNSS" refers to the USA's Global Positioning System (GPS), Russia's GLONASS, Europe's GALILEO and China's BEIDOU navigation satellite system (BDS). It is noted that GPS, GLONASS, Galileo and BDS use different reference time systems creating time offsets between them. However, the offsets can be determined at the system level or user level. Any one or more constellations can offer the same global standard UTC time. With their worldwide coverage, continuous service, GNSS has become one of the most efficient and standard systems for time dissemination in many applications. Many industries such as energy, meteorology and telecommunications rely on GNSS for accurate time synchronization in their systems and devices. The accuracy achieved by GNSS-based time synchronization using standard GPS PNT service is better than 40 ns 95% of time [62]. This can meet the most stringent requirements for VANET time synchronization.

This section begins with discussing the motivation for using GNSS for VANET time synchronization. This is followed by descriptions of GNSS models and methods for time transfer. The challenges and solutions due to absence of GNSS signals in vehicular environments are then summarised.

### 6.1. Motivation of GNSS-driven time synchronization in VANETs

Most of the earth-based time transfer techniques suffer from path delay measurement uncertainties. In contrast, the satellite-based GNSS time transfer systems possess measurable constant path delays. This arises because the variation of path delays are small and due to clear, unobstructed paths to receivers. Therefore, the delay measurements are straightforward and can be more easily calibrated compared to any ground-based systems. In addition, the radio interferences due to weather or any other ground-based noise have less impact in satellite-based GNSS systems.

In telecommunication networks, GNSS is used to synchronize some major nodes called root or server nodes outdoors. Through these root nodes, other nodes in the system are synchronized by using other synchronization techniques, which are mostly based on message transfer between nodes.

In contrast to telecommunication networks, VANETs are outdoor-based networks. Except in tunnels and some signal blocked roads, nodes in VANETs on the road are mostly under the coverage of GNSS signals. It is a straightforward choice for VANETs to use GNSS for synchronization. GNSS receivers have already been used for vehicle navigation and positioning. Nowadays, multi-GNSS constellations, more precise GNSS services, such as space-based argumentation systems (SBAS), differential GNSS (DGNSS) services and precise point positioning (PPP), are available for VANET deployments. The GNSS-based time synchronization is indeed plausible in VANETs.

It is therefore prudent to understand how GNSS time solutions provide synchronization in VANETs and what the possible solutions are when GNSS services are absent, such as when vehicles travel in tunnels. The feasibility and accuracy of GNSS time solutions are investigated in recent studies [62,63].

### 6.2. GNSS time synchronization models for VANET

Different GNSS systems follow the same estimation principle for position, velocity and time computing. Without loss of generality, this section discusses the theory of GPS time, time transfer from GPS, and time propagation. It explains time synchronization model by using GPS data. It also outlines possible support of synchronization in the absence of GPS signals.

#### 6.2.1. GPS time

GPS time is one of the standard times related to UTC. It is a continuous time generated from a precise atomic clock and maintained by the GPS control segment. GPS time is related to UTC by leap seconds. At present, GPS time is 18 s ahead of UTC time. This means that the leap seconds between GPS time and UTC time are 18 s. This is indicated in USNO navy's website tycho.usno.navy.mil/leapsec.html.

#### 6.2.2. GPS time receiver and time transfer

There are a variety of GPS receivers which differ within applications, technologies and manufacturers. Most consumer-grade GPS receivers receive single-frequency C/A code. The clocks in GPS receivers are mostly quartz clocks. They are synchronized by GPS signals. The GPS receiver clock solution is obtained from the pseudo-range measurements. For the purpose of this work, the pseudo-range measurement can be written as [64]:

$$P_u = \rho + cdt + \varepsilon \qquad (12)$$

where $P_u$ is the Pseudo-range measurement, $\rho$ is the geometric distance between the receiver and the satellite, $c$ is the speed of light, $dt$ is the receiver time offset or bias with respect to the receive clock time tag, and $\varepsilon$ is the sum of all errors.

It is clear from Equation (12) that $dt$ can be directly obtained from the observed-computed (O-C) difference ($P_u - \rho$) if the distance is known. Taking average or weighted average over all the O-C differences would improve the accuracy of the $dt$ solutions. This is the static mode for time transfer. However in a VANET, the vehicle nodes are moving. The distances $\rho$ are computed with the approximate vehicle states $X_0$. The coordinate biases of $X_0$ can affect the accuracy of dt solution if ignored. Considering the coordinate biases, Equation (12) can be rewritten as:

$$P_u - \rho(X_0) = \frac{\partial \rho_0}{\partial X_0} dX + cdt + \varepsilon \qquad (13)$$

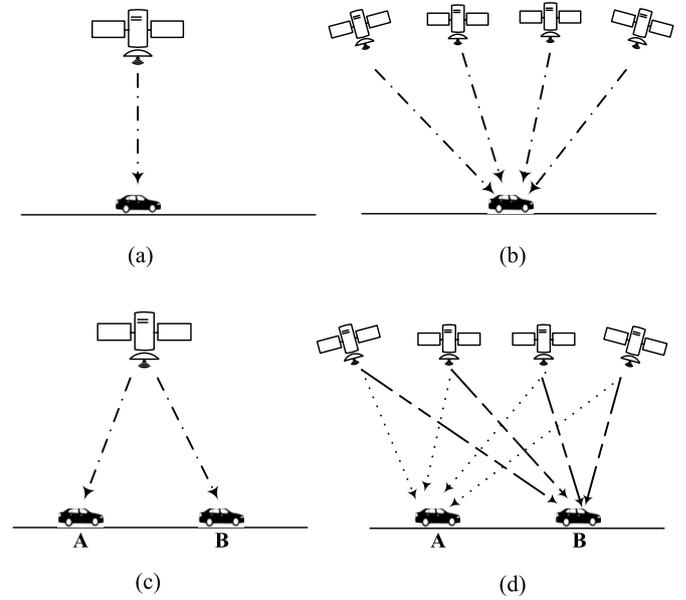

**Fig. 10.** GPS time transfer.

where the partial derivatives of the geometric distance $\rho_0$ are computed with respect to the 3-dimensional approximate coordinate vector $X_0$; $dX$ is the 3-dimensional position deviation with respect to the approximate states $X_0$. They can be estimated along with the clock bias dt. The least square or weighted least square procedures are usually applied to solve the estimation problem with four or more satellites in view [65]. Time solution may also be obtained with as few as one satellite, since $dX$ is often determinable by alternative onboard positioning sensors to a degraded accuracy.

Several techniques have been developed to transfer GPS time based on the above equation (13). The simplest method is 'time dissemination', which is also known as 'One Way' method. It predominantly aims to synchronize an on-time pulse, or to calibrate a clock frequency source. Fig. 10(a) illustrates the one-way concept where the clock bias is determined by the difference between observed range $P_u$ and computed range $\rho$, namely (O-C). With more satellites in view as shown in Fig. 10(b), the clock bias can be estimated from the average of all the (O-C)s. The user-position biases will affect the clock bias dt solutions. As long as four or more satellites in view the coordinate bias vector $dX$ and clock bias $dt$ can be determined with the linear Equation (13). A typical clock solution accuracy with GPS-only signals is 40 ns [62].

A more accurate and elegant technique for GPS time transfer is 'Common View'. As shown in Fig. 10(c), it measures the clock bias difference between two receiver oscillators using the difference of the (O-C)s between two receivers. This differencing leads to the cancellation of satellite orbit and clock error and local ionosphere and troposphere delays, thus providing a higher accuracy for the clock offset, saying in the level of 10 ns. Similarly when multiple satellites are in view as shown in Fig. 10(d), the common view method is equivalent to differential GPS, and determines the 3D coordinate and clock offsets between two receivers.

There is a highly accurate technique for GPS time transfer called the 'Carrier-Phase' method. In this method, both L1 and L2 carrier phase signals are used to calculate time [66]. The timing accuracy achieved from this method is in sub-nanosecond level. However, dual-frequency phase receivers are more costly and may be not a popular choice for vehicle users.

#### 6.2.3. A simple model for GNSS time synchronization

Based on the above "one-way" and "common-view" modes, the GNSS time synchronization model is outlined as follows. An on-board GPS receiver tracks satellites. Once the clock bias dt is obtained with a one-way time transfer, the receiver can determine its UTC time $t_{UTC}$,

$$t_{UTC} = t_u - dt - dt_{UTC} \quad (14)$$

where $t_u$ is the receiver time, which is usually the time tag of a standard time epoch; $dt_{UTC}$ is the offset between GPS time and UTC time, which includes a integer term for leap seconds and a fractional correction term calculated from GPS navigation messages. Both are the same for different receivers at the same time. In the common view mode, receiver B obtains the clock bias with respect receiver A, i.e., $dt_{BA}$, the receiver B's UTC time is obtained as follows:

$$t_{B_{UTC}} = t_{A_{UTC}} - dt_{BA} \quad (15)$$

A typical GNSS receiver has an internal quartz-based oscillator that continuously runs and follows GPS time. Generally, in the one-way time transfer, the clock update rate can be the same as the receiver sample rate. A low-end receiver updates its outputs at 5 to 10 Hz, while a high-end geodetic receiver's sample rates can be up to 50 Hz. However, such quartz clocks still exhibits deviations because the frequency of each clock is different and tend to diverge from each other. This divergence is known as clock skew. The clock drifts with respect to time is the derivative of clock skew [13]. Following [67] and [68], in general, a node clock in a GNSS-synchronized distributed network over a time interval of minutes to hours can be characterised as:

$$C_i(t) = d_i.t + b_i \quad (16)$$

where $t$ is the time corresponding to the UTC time. $d_i$ is the clock drift due to the oscillator's frequency differences and the result of to the environmental changes at the node, e.g., variations in temperature, pressure and power supply voltage. $b_i$ is the offset between the receiver local clock and the UTC time obtained from one-way time transfer approach. This reflects the effect of hardware delays of the clock.

Any two such GNSS-synchronized clocks can be represented as:

$$\left.\begin{array}{l} C_1(t) = d_1.t + b_1 \\ C_2(t) = d_2.t + b_2 \end{array}\right\} \quad (17)$$

They can also be related as follows:

$$C_1(t) = \Theta_{12} C_2(t) + \beta_{12} \quad (18)$$

where $\Theta_{12}$ is the relative drift between two receivers and $b_{12}$ is the offsets due to the bias variations. If the two receivers are the same model, the relative offset can be small and the drift $\Theta_{12}$ is 1.

To measure the time offset between two GNSS powered nodes, experiments have been performed at physical layer on 1PPS signals generated by two GPS receivers. The experimental results are presented in [62,69]. It has been shown that two same-model consumer-grade GNSS receivers are capable of synchronizing network nodes with nano-second scale timing accuracy.

### 6.3. Challenges and solutions in the absence of GNSS signals

Signal transmissions between satellites and receivers solely relies on the principle of the Line-of-Sight (LOS) wave propagation technique. Drivers often experience outages of navigations when driving through high-rise streets. This does not necessarily mean loss of time synchronization. First, GNSS time solutions can be obtained with a single satellite at reduced accuracy. The availability of valid time solutions is much higher than the availability of valid position solution. For instance, a vehicle experiment shows that the percentage of valid GPS+Beidou position solutions over some Brisbane high-rise streets is 80.25%, while the percentage of a minimum of one satellite is 100% [62]. As mentioned, time solution can still be available with a single satellite, despite having a degraded accuracy.

Secondly, some measures have been proposed as fall back solutions in the blockages of GPS signals. One of them is to switch from normal mode to holdover mode using GPS Disciplined oscillator (GPSDO). GPSDO is a specially made firmware for holdover mode. It enables the internal oscillator to predict and imitate the original timing and frequency of the GNSS system. A GPSDO is primarily made of a phase detector and voltage control oscillator (VCO). Its fundamental purpose is to acquire information from the GNSS signal of satellites to control the frequency of local quartz or rubidium oscillators. When GPS signals are unavailable, GPSDO keeps its oscillation in a stable frequency using the knowledge of its past performance.

In order to boost the performance of GPSDO, some additional technique have been developed such as adaptive temperature and aging compensation during the holdover period [70]. The adaptive temperature and aging compensation are based on a recursive implementation of linear regression. As an additional circuitry, a simple semiconductor ambient temperature sensor and an A/D converter are used. The performance of both types improved GPSDOs are the same to some extent. However, it is not well defined how long the independent free-running GPSDOs are executed. Nevertheless, experiments have been conducted to test the performance of GPSDO. An experiment was carried out over a week for holdover on 4 GPSDOs, in which an oscillator is made of quartz and the other three are made of rubidium [35,71]. After a week, the time offset from the quartz oscillator was shown to be 82 μs. In comparison, the best time offset performance of less than 3 μs was measured for the three rubidium oscillators. This level of synchronization accuracy is considered to be acceptable for VANET time synchronization applications.

Finally, the problem of GPS signal blockages can be addressed by incorporating GPS synchronization with other methods. If some vehicles nodes that can view satellites have GNSS time solutions, they act as root servers for synchronization of other nodes through a non-GPS time synchronization technique. NTP-GPS is the backbone of general computer networks, in which the standard time hosting servers are synchronized with GPS. For example, time synchronization based on absolute GPS is employed in Automatic Identification System (AIS) for ships [72]. In DSRC-based networks, Time advertisement (TA) has been specified in the IEEE1609.4 to provide time solutions to other devices where GNSS signals are not available. However, to date the performance of TA for VANET has not been well understood.

## 7. Concluding remarks

Communications in VANETs involve V2V and V2I communications. They form the basis of VANETs for network connectivity and various safety applications on roads. Due to the highly dynamic and mobile characteristics, precise timing and accurate measurement of transmission delay become critical in VANETs. Time synchronization helps establish an agreed time over VANETs. It enables proper coordination and consistency of various events throughout the networks. It also allows accurate sequencing and real-time control of message exchanges over the networks.

This paper has investigated why time synchronization is necessary and what the time synchronization requirements are for VANETs. It has shown that most existing synchronization techniques used for other types of wireless networks are not directly applicable to VANETs. The discussions are accompanied by detailed evaluations of existing time synchronization protocols in various distributed network systems.

Maintaining time synchronization is challenging in VANETs. In some vehicular applications, VANETs require highly accurate time synchronization. Some security measures also need precise time synchronization, which is currently not achievable in VANET environments. Synchronization techniques developed for general WSNs could face compatibility issues if they were applied to VANETs. GNSS-driven time synchronization is a promising technique for VANETs. The impact of blockage of GPS signals on timing is less serious than the impact on positioning, but it is still a problem to be addressed by integrating other synchronization methods.

In summary, the paper has investigated time synchronization in VANETs from both the research and development perspectives. It has covered the basic theory and analysis of existing synchronization protocols as well as comparisons of these protocols. The paper has also highlighted the advantages of GNSS in VANET time synchronization. Further effort in research and development is required to make GNSS-driven time synchronization practical for VANETs.

## Acknowledgement

This work is supported in part by the Australian Research Council (ARC) under the Discovery Projects Scheme (grant Nos. DP160102571 and DP170103305) to Author Y.-C. Tian.